# Micromagnetic study of electrical-field-assisted magnetization switching in MTJ devices

M. Carpentieri[1], R. Tomasello[2], M. Ricci[3], P. Burrascano[3], G. Finocchio[4]

[1]Department of Electrical and Information Engineering, Politecnico of Bari, via E. Orabona 4, I-70125 Bari, Italy
[2]Department of Computer Science, Modelling, Electronics and System Science, University of Calabria, Rende (CS), Italy
[3]Department of Engineering, Polo Scientifico Didattico di Terni, University of Perugia, Terni, TR, I-50100 Italy
[4]Department of Electronic Engineering, Industrial Chemistry and Engineering, University of Messina, Messina, Italy

Perpendicular MgO-based Magnetic Tunnel Junctions are optimal candidates as building block of Spin Transfer Torque (STT) magnetoresistive memories. However, up to now, the only STT is not enough to achieve switching current density below $10^6$ A/cm². A recent work [Wang *et al.*, *Nature Mater.*, vol. 11, pp 64-68, Jan. 2012] has experimentally demonstrated the possibility to perform magnetization switching assisted by an electric-field at ultra-low current density. Theoretically, this switching has been studied by using a macrospin approach only. Here, we show a full micromagnetic study. We found that the switching occurs via a complex nucleation process including the nucleation of magnetic vortexes.

## I. INTRODUCTION

MAGNETIC Tunnel Junctions (MTJs) with a ferromagnetic CoFeB free layer and an MgO spacer, have excellent characteristics to be used in Spin Transfer Torque Magnetoresistive Random Access Memories (STT-MRAM) [1], [2]. The magnetization reversal induced by spin torque was firstly observed in in-plane structures [3], [4] and, therefore, it was successfully achieved in perpendicular devices [5], [6]. Reductions of the critical current and high thermal stability have been obtained with different strategies [7], [8], [9], [10], [11], [12] on top of those the use of interfacial perpendicular anisotropy arising at the interface between Fe-rich CoFeB and MgO [13], [14], [15]. However, the reversal current density is still too high, hence, besides the only STT, other switching mechanisms have to be investigated. Many experimental [16], [17], and macrospin [18] studies have addressed the magnetization reversal assisted by the voltage controlled magnetocrystalline anisotropy (VCMA) [19] and, among them, a milestone experiment was performed by Wang *et al.* [20]. They have demonstrated that switching processes can be controlled by a trade-off between a bias field $H_{ext}$ and an unipolar voltage pulse, achieving the magnetization switching for current densities of the order of $10^4$ A/cm² in presence of a small bias field $H_{ext}$ (+z direction)For the initial parallel state (both magnetization along -z direction), a low voltage pulse is applied to decreases the perpendicular anisotropy in order that the current flowing through the stack switch the state from the parallel (P) to antiparallel (AP). The P→AP switching is assisted by the external field. For AP-->P switching the scenario is different. A high voltage pulse reduces drastically the perpendicular anisotropy, leading to an in-plane free layer magnetization. In this case, the STT related to the large current density flowing through the MTJ stack switch the magnetization. Here, we perform a numerical study by means of micromagnetic simulations to deeply understand that experiment[20]. Our results predict the achievement of electrical-assisted-magnetization-reversal in the nanosecond scale at $J=10^5$A/cm², making such switching mechanism more suitable for high speed storage applications. The switching occurs via the nucleation of complex magnetization patterns, including vortex and antivortex and, therefore, it is strongly spatially non-uniform, making impossible its understanding in the framework of macrospin approximation.

The paper is organized as follows. Section II introduces the device and the numerical details. Section III describes the achieved results and explains the physical origin of the switching process. Conclusions are reported in Section IV.

## II. NUMERICAL MODEL

We study an MTJ stack composed by CoFeB(1.3nm)\MgO(1.2nm) \CoFeB(1.6nm). The thinner and the thicker layers act as pinned and free layer respectively. A sketch of the structure is represented in Fig. 1. We introduce a Cartesian coordinate system with the *z*-axis oriented along the out-of-plane direction and the *x*- and *y*-axis positioned into the film plane. The device has a circular cross section with a diameter $D$=200 nm, a free layer saturation magnetization $M_s$=9x10⁵ A/m and an exchange constant $A$=2x10⁻¹¹ J/m. Since the free layer is very thin and it is coupled to a MgO insulator, we can consider a VCMA $K_u$=5.3x10⁵ J/m³ [13], which is large enough to stabilize an out-of-plane magnetization easy axis. The magnetization of the pinned layer points down (negative *z*–direction) and an external magnetic field $H_{ext}$=5.5 mT is applied along the positive *z*-direction. Initially, the free and pinned layer magnetizations are collinear.

The voltage dependence of the anisotropy constant is linear as shown in [20]. The anisotropy constant is reduced to $K_u$=4.8x10⁵ J/m³ for the current pulse to switch P→AP while its value is to 4.4x10⁵ J/m³ for the current pulse to switch AP→P.

**FIG. 1 HERE**

The micromagnetic simulations are based on the numerical solution of the Landau-Lifshitz-Gilbert-Slonczewski equation:

$$(1+\alpha^2)\frac{d\mathbf{m}}{dt} = -(\mathbf{m}\times\mathbf{h}_{eff}) - \alpha\mathbf{m}\times(\mathbf{m}\times\mathbf{h}_{eff})$$
$$- \frac{g|\mu_B|J_{MTJ}}{e\gamma_0 M_S^2 d_{FL}}\varepsilon(\mathbf{m},\mathbf{m}_p)[\mathbf{m}\times(\mathbf{m}\times\mathbf{m}_P) - q(V)(\mathbf{m}\times\mathbf{m}_P)] \quad (1)$$

where $\alpha$ is the Gilbert damping, $\mathbf{m}$ is the magnetization vector, $\mathbf{h}_{eff}$ is the effective field, $g$ is the Landè factor, $\mu_B$ is the Bohr magneton, $e$ is the electron charge, $\gamma_0$ is the gyromagnetic ratio, $M_S$ is the free layer saturation magnetization, $d_{FL}$ is the thickness of the free layer, $J_{MTJ}$ is the current density flowing perpendicular to the structure, $\varepsilon(\mathbf{m},\mathbf{m}_p)$ characterizes the angular dependence of the Slonczewski spin torque term and $\mathbf{m}_P$ is the magnetization of the pinned layer. $q(V)$ is the voltage dependent parameter for the perpendicular torque, where $V$ is the voltage computed from the bias voltage dependence of the TMR and the current density. We set $\alpha$=0.015, a typical value for this material [13]. The thermal field $\mathbf{h}_{th}$, which is included as an additive term to the effective field, is a random fluctuating three-dimensional vector quantity, given by $\mathbf{h}_{th} = (\xi/M_s)\sqrt{2(\alpha K_B T / \mu_0 \gamma_0 \Delta V M_s \Delta t)}$, where $K_B$ is the Boltzmann constant, $\Delta V$ is the volume of the computational cubic cell, $\Delta t$ is the simulation time step, $T$ is the temperature of the sample [21], and $\xi$ is a random number from a Gaussian distribution with zero mean and unit variance [22]. Detailed numerical description of the model and the algorithms can be found in [23], [24].

### III. RESULTS AND DISCUSSION

Fig. 2a shows the current pulses (as consequence of voltage pulses) as function of time $t$. We consider that the switching phenomenon is achieved when the normalized $z$-component of the magnetization reaches $\pm 0.8$. The voltage pulses give rise to an anisotropy constant reduction and a current flow through the MTJ. The first pulse, long 15 ns, gives rise to a negative current density of about $6\times 10^5$ A/cm$^2$. During this time range, the $z$-component of the magnetization starts to reverse (see Fig. 2b), remaining in an oscillatory state. At $t$=18 ns, when the first current pulse is switched off, $K_u$ is turned back to the initial value, and due to the external field $H_{ext}$ (pointing upwards), the P→AP switching is fully attained. This is a stable state and the <$m_z$> keeps firmly the AP state. At $t$=38 ns the second higher pulse is applied ($J_{MTJ}$=-7x10$^5$ A/cm$^2$) for a time of 3 ns. Consequently, since $K_u$ reduces more and the negative current favors the P state, the AP state becomes unstable, accordingly <$m_z$> starts to switch back. When the current pulse is switched off ($t$=41 ns), the anisotropy field returns to the initial value and helps the full magnetization reversal. At $t$=47 ns, the AP→P switching is achieved. The current densities to achieve both switching processes in less than 50 ns are about one order of magnitude smaller than the one used in STT-MRAM [14].

**FIG. 2 HERE**

The P→AP is approximately insensitive to the pulse time duration. With this in mind, we systematically study the AP→P switching as function of the duration of the current pulse. In detail, we sweep the pulse duration from 2.6 to 7 ns, with a time step of 0.2 ns, finding that only when the pulse time is 2.8, 3 and 5.6 ns the switching is obtained. In Fig. 3, we illustrate the temporal evolution of the three components <$m_x$>, <$m_y$> and <$m_z$> of the magnetization when the pulse duration is 3 ns (Fig. 3a) and when it is 2.6 ns (Fig. 3b). After the current pulse is switched off, in the former case <$m_z$> switches from AP to P, whereas in the latter one the reverse does not occur and <$m_z$> returns to the AP state (<$m_z$> =+1).

**FIG. 3 HERE**

The origin of such dependence can be understood by studying the spatial configuration of the magnetization before and after the current pulse is removed. Fig. 4 represents the spatial distribution of the magnetization in the case where the reversal occurs (Fig. 4a and 4b) and when it is not achieved (Fig. 4c and 4d). In out-of-plane MgO-based MTJs, the easy-axis of the magnetization is mainly dependent on the competition between the demagnetizing field $\mathbf{h}_d$ and the perpendicular anisotropy field $\mathbf{h}_{K\perp}$. Particularly, the $z$-component of $\mathbf{h}_d$ tries to lead the magnetization into the free layer plane, whereas $\mathbf{h}_{K\perp}$ keeps the magnetization in the out-of-plane direction. Once we reduce the perpendicular anisotropy by applying the current pulse, $\mathbf{h}_d$ becomes comparable with $\mathbf{h}_{K\perp}$, and, therefore, we obtain a quasi-in-plane structure. In this framework, the applied current is able to excite non-uniform chiral magnetization configurations such as vortexes and antivortexes [25].

**FIG. 4 HERE**

When the current pulse is applied, the circular symmetry of the Oersted field nucleates a vortex state. In this case, two vortexes are generated and they rotate in the plane of the structure. Particularly, in Fig. 4a and 4b we can clearly identify two vortexes and two antivortexes (indicated with "V" and "AV" respectively in the figures); on the other hand, in Fig. 4c and 4d, the second antivortex is pushed out from the free layer and only one antivortex is evident together with two vortexes. In addition, in the cases when the switching is obtained, a larger portion of the magnetization points down along the negative $z$-axis (Fig. 4a and 4b show blue background color), with respect to the other two cases (Fig. 4c and 4d). After removing the current pulse, the spatial distribution of the magnetization evolves in a central domain oriented in the positive out-of-plane direction and in two side domains pointing along $-z$-direction (see Fig. 4e). At a later stage, the two side domains collapse each other forming only one domain. Finally, it expands across the whole section of the free layer. A magnetization configuration similar to the previous case is also observed when the current pulse is long 2.6 ns (Fig. 4f), with the exception that the central domain is larger and the lateral domains smaller. Subsequently, the central domain expands, bringing back the magnetization in the AP state. After removing the current pulse, the vortex state (with +z polarity) is destroyed relaxing into a more uniform magnetization configuration, pointing upwards. The width of this region depends on the vortexes spatial position before removing the pulse. If the area of the nucleated central domain is confined in the centre of the free layer, the side domains (pointing downwards) are quite large to favour energetically the –z magnetization direction (case of Fig. 4e).

Just to summarize, the AP→P switching is achieved only in the case where two vortexes and two antivortexes are simultaneously nucleated just before the current pulse is switched off. On the other hand, when two vortexes and only one antivortex are generated, the AP→P switching does not occur. The number of chiral configurations (four or three) excited when the current pulse is removed depends on the pulse duration.

Finally, we introduce the thermal field as computed at room temperature $T$=300 K. Figure 5 shows the $<m_z>$ behavior for different current pulse durations in the P→AP switching (Fig. 5a) and in the AP→P one (Fig. 5b). In the first reversal process, pulse durations from 6 ns to 15 ns are taken into account and in all the cases the magnetization is fully reversed within 30 ns. On the other hand, the thermal effects make the second switching harder. In fact, only when the pulse durations are 3 and 6.2 ns the magnetization switching is obtained and for time larger than 80 ns. Therefore, the thermal effect triggers the P→AP switching and delays the process for the AP→P one. This is related to the fact that the P→AP process is strongly dependent on the external field, while the AP→P is mainly related to STT.

## IV. Conclusion

In summary, the switching assisted by an electric field, studied experimentally in [20], has been described within a micromagnetic framework. The dependence of the entire switching process on the current pulse time duration has been analyzed, highlighting that the AP→P one occurs only when two vortexes and two antivortexes are simultaneously excited before the current pulse is switched off. Thus, the switching behavior is characterized by strongly non-uniform magnetization configurations and, for this reason, it is not possible to describe it by using a simple macrospin model. Furthermore, with respect to the experimental case, it has been shown how the whole magnetization reversal process can be executed in a time of about 50 ns, and, by considering also the thermal effect at room temperature, within about 100 ns (much less than the experimental evidence). In this perspective, the results suggest that the electrical-field-assisted switching mechanism can be very useful and achievable from a technological point of view.

## Acknowledgment

This work was supported by the project PRIN2010ECA8P3 from Italian MIUR. MR and PB acknowledge financial support from Fondazione CARIT. MC acknowledges financial support by the project "RES NOVAE - Reti, Edifici, Strade - Nuovi Obiettivi Virtuosi per l'Ambiente e l'Energia". This project is supported by the Italian University and Research National Ministry research and competitiveness program that Italy is developing to promote "Smart Cities Communities and Social Innovation".

HQ-10                                                                                                                 4

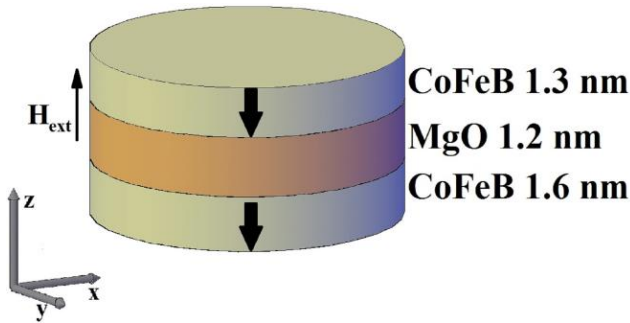

Fig.1. Sketch of the studied MTJ device.

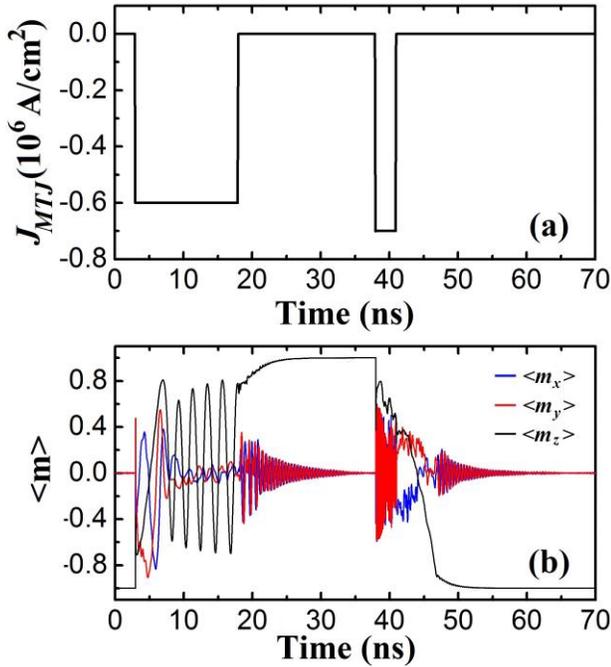

Fig. 2. (a) Applied current pulses. (b) Time domain plot of the three normalized components of the magnetization ($<m_x>$, $<m_y>$ and $<m_z>$, respectively represented in blue, red and black) during a whole reversal process (P→AP and AP→P) as a function of the current pulses in (a).

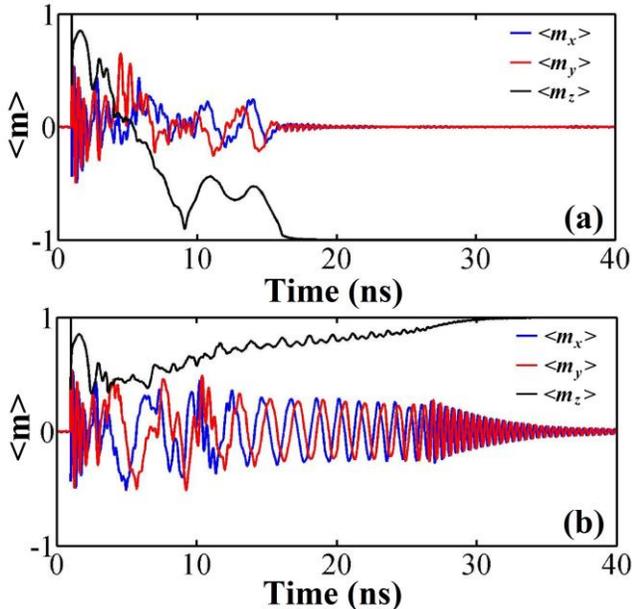

Fig. 3. Time domain plot for the three normalized components of the magnetization $<m_x>$, $<m_y>$ and $<m_z>$, respectively represented in blue, red and black. (a) Pulse time of 3 ns. (b) Pulse time of 2.6 ns.

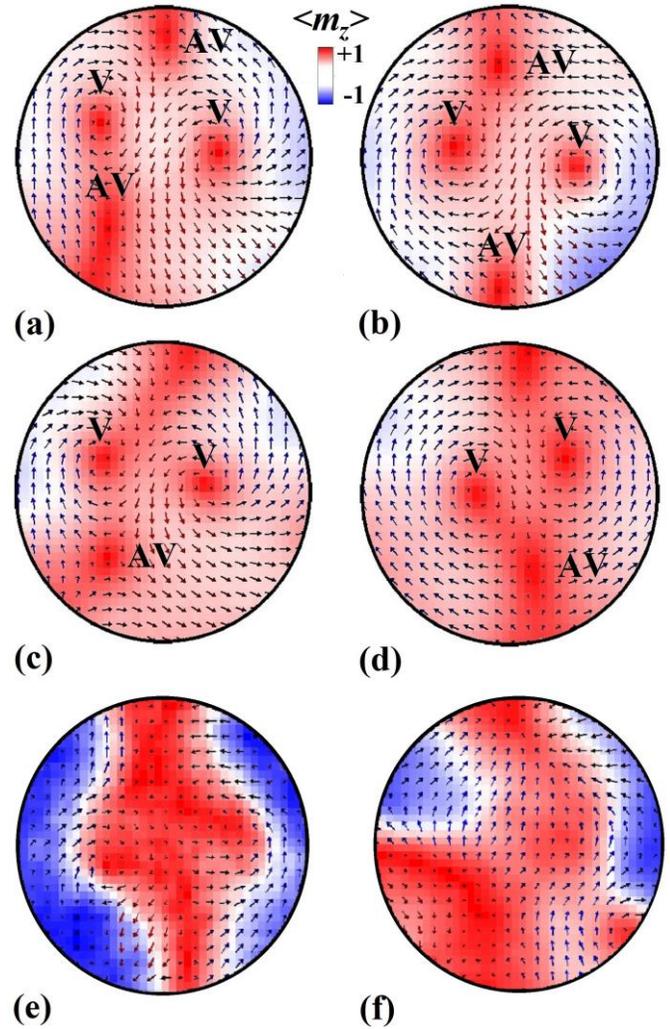

Fig. 4. Spatial magnetization configurations before the current pulse is switched off. The arrows refer to the y-component of the magnetization (blue positive and red negative), whereas the background colors refer to the z-component (red positive direction and blue negative direction). The letter "V" indicates vortexes, while "AV" refers to antivortexes. (a) Pulse time of 3 ns. (b) Pulse time of 2.8 ns. (c) Pulse time of 2.6 ns. (d) Pulse time of 3.4 ns. (e) Spatial magnetization configuration after the current pulse is removed for a pulse time (e) $t=3$ ns and (f) $t=2.6$ ns.

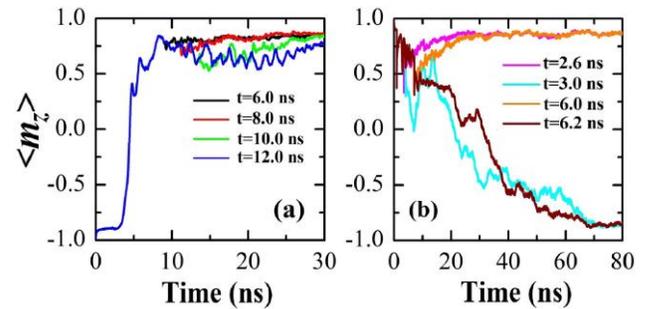

Fig. 5. Time domain plot for the z-component of the magnetization for different time pulses at $T=300$ K. (a) P→AP switching process. (b) AP→P switching process.